\documentclass[aps,amsmath,twocolumn]{revtex4}
\usepackage{graphicx}

\begin{document}


\title{The CMB flexes its BICEPs while walking the \textit{Planck}}

\author{Douglas Scott} \email{docslugtoast@phas.ubc.ca}
\author{Ali Frolop} \email{afrolop@phas.ubc.ca}
\affiliation{Department of Physics \& Astronomy\\
University of British Columbia,
Vancouver, BC, V6T 1Z1  Canada}

\date{1st April 2014}

\begin{abstract}
Recent microwave polarization measurements from the BICEP2 experiment may
reveal a long-sought signature of inflation.  However, these new results appear
inconsistent with the best-fit model from the {\it Planck\/} satellite.  We
suggest a particularly simple idea for reconciling these data-sets, and for
explaining a wide range of phenomena on the cosmic microwave sky.
\end{abstract}


\maketitle

\date{today}

\noindent
\paragraph*{Background}
As measurements of the cosmic microwave background (CMB) anisotropies
become more precise, and the standard model of cosmology continues to be a
good fit, the attention of the community has shifted onto the much more
important business of explaining minor tensions and anomalies.  A case in
point arises from recent results from the BICEP2 experiment \cite{BICEP2},
which appear discrepant with the best-fitting model from the
{\it Planck\/} satellite \cite{PlanckXVI}.

BICEP2 has achieved the deepest observations yet of CMB polarization.
These measurements have revealed a particular signature in the polarization
that is quite hard to explain -- sometimes called `B-modes', but more
technically known as `a swirly pattern'.
These results are certainly exciting \cite{Frolop},
since they have implications for a parameter called $r$.
On learning about these new results, the main question that people have is
`what is $r$?'  The answer is that
$r$ equals 0.2 from BICEP2.  However, this is seen as somewhat problematic,
since the upper limit from {\it Planck\/} is 0.11.

There are many ways in which one could resolve this apparent discrepancy.
Here we wish to focus on the simplest explanation, which also solves at
least two additional anomalies in CMB data.  We will later describe how this
explanation has a natural foundation in a radical reinterpretation of the
CMB sky.

\noindent\paragraph*{Perturbations}
It is fairly well established that the Universe contains lumps \cite{lumps}.
These lumps can either be in a form that come in a wide range of scales (called
`scalars') and those which help resolve tensions in the data (called
`tensors').  The tensors that are the hardest to explain are gravitational
waves.  The quantity $r$ has something to do with the amount of these
gravitational waves compared with other kinds of lumps \cite{ratio}.

Among the anomalies seen in data from the {\sl WMAP\/} satellite
\cite{Hinshaw03} and confirmed by {\it Planck\/} \cite{PlanckI}, the most
striking are: the fact that the power in
the lowest multipoles seems to be low (the so-called
`low-low-ells' \cite{PlanckXVI}); and
the fact that on relatively large angular scales, one side of the sky contains
more power than the other side
(sometimes called the `lopsided Universe' \cite{PlanckXXIII}).

Gravity waves only affect the largest angular scales in the CMB \cite{TWL},
corresponding closely with the multipoles that appear to be affected by
the `lopsidedness'.  Hence we can explain why one side of the Universe has
more anisotropy power by simply having more gravity waves on one side than
the other.  The lowness of the low multipoles in the overall power spectrum
is then explained by the masking of our Galaxy, which favours the side of
the sky with lower tensors.  And the discrepancy between BICEP2 and
{\it Planck\/} disappears when one assumes that the small part of the sky
mapped by BICEP2 lies in the direction with greatest tensors \cite{But}.
Hence one just needs to take a universe with a gradient in the
gravity waves, make that gradient maximal (surely the most obvious amplitude
to choose) in the right direction, and leave the usual kinds of lumps alone
-- then one can solve at least three cosmological problems at once.
With a little effort it seems likely that this idea could also explain other
anomalies, such as the Cold Spot and Galactic Haze, as well as resolving
discrepancies in $H_0$, $\sigma_8$, and other cosmological parameters.

\noindent\paragraph*{Model}
Is there an underlying physical model that could explain such a
universe?  To answer this, we need to take a little digression.
In order to describe the CMB last-scattering surface, and the physics of the
acoustic modes, cosmologists often use an analogy with the photosphere of a
star \cite{photosphere}.
We see {\it out\/} into the `cosmic photosphere' in all directions around us,
just as we only see {\it into\/} the optical depth ${\sim}\,1$ surface when
we look at a star like the Sun.  Hence the last-scattering
surface surrounding us is just like an inside-out star.  Helioseismologists and
CMBologists both use spherical harmonics to describe surface variations, so
that they can probe the physics of the interior of the hidden acoustic cavity
\cite{CMS}.  The analogy is an extremely close one, with, for example, the
angular scale of damping and the main acoustic feature for the CMB
corresponding, respectively, to granulation and super-granulation for the Sun.

In fact this correspondence is so good that it suggests there is much
more than an analogy going on here.  We therefore suggest that the apparent
equivalence is actually real.  We can achieve the `inside-out-ness'
through the use of a particular conformal transformation, which takes large
scales into small scales, and vice versa.  This is probably related to the
`Conformal Cyclic Cosmology' \cite{Penrose}, just without the `cyclic' part.
It is also related to the idea of `duality' in string theory.  We can
imagine that our CMB sky is effectively the same as the surface of a star
like the Sun in some other dimension.  In other words we have a
relationship between Another-dimension's Sun and the CMB, which we could
call the AdS-CMB correspondence.

The full transformations require that the
entirety of space-time be projected onto the bounding surface, and hence it is
consistent with holographic conjectures for our Universe.  The conformal
invariance of the Weyl projection operators across the boundary mean that
tachyonic fields suffer a conjugation, leading to a reciprocal Lyapunov
entropy \cite{probably}.  This means that rapid variations on the surface of
the dual Sun correspond to long timescales on our CMB sky, explaining why
the CMB takes so long to evolve \cite{varyingCMB}.

With the CMB being dual to a star, we can easily understand the other
basic properties of our microwave sky.  Gravitational ripples in the fabric
of our space-time are simply gravity waves in the AdS \cite{GW}.
One side of the sky has more of these waves because that direction lies
at the bottom of the star.  The polarization
pattern develops a handedness just like the pattern seen around sun-spots, with
the B-modes on our sky corresponding to actual ${\vec B}$-fields on the surface
of the dual star.  In fact all of the phenomenology of the CMB becomes much
clearer when we realise that it is simply the same as the physics on the
surface of a conformally-transformed star.

Perhaps observers in this other reality are
looking at a star right now, with our Universe contained inside the
stellar photosphere.  The idea of the `multiverse' \cite{Moorcock}
is also clear from this
perspective, since presumably there are many different stars.

\noindent\paragraph*{Conclusion}
We feel confident that with further development our idea will be able to
explain many other mysteries.  It is only a matter of applying some
imagination to explain what came before the Big Bang, the origin of
primordial magnetic fields, ultra-high energy cosmic rays, the cuspy halos
problem, the emergence of the dark energy, the arrow of time,
baryogenesis and Saturn's polar hexagon,
as well as results from PAMELA, DAMA, OPERA, Pioneer, and more \cite{us}.

All of this makes complete sense if we simply consider the entire Universe to
be turned inside-out, as well as inverted in scale and time.  The only
alternative being discussed in the literature would be to suppose that
experimental results might shift slightly, so that the {\it Planck}, BICEP2 and
other cosmological data-sets become consistent with the so-called
$V\propto m^2\phi^2$
inflationary potential.  However, that would imply that the effective
potential in the early Universe had a minimum with a quadratic shape.
Such a shape would be unprecedented in physics, and hence that
solution seems quite preposterous compared with ours \cite{further}.


\smallskip



\begin{thebibliography}{9}
\bibitem{lumps} Some of our best friends are lumps.
\bibitem{ratio} It would have been more convenient if $r$ had been defined
 as simply the ratio between tensors and scalars.
\bibitem{But} The rather inconvenient details (namely that none of the
 directions are very close to what would be required), haven't been allowed
 to get in the way of a good story here.
\bibitem{photosphere} The origin of this analogy is uncertain.  A poll of our
senior colleagues resulted, unsurprisingly, in most of them claiming credit
for themselves.
\bibitem{GW} Hence when pedantic colleagues insist that `gravitational waves'
are not the same as `gravity waves', we can just smile knowingly.
\bibitem{further} Further such suggestions should be sent to the second
 author, Dr.~Frolop.
\end{thebibliography}
\end{document}